\documentclass[onecolumn,showpacs,amsfonts,aps,prc,nofootinbib,floatfix,%
superscriptaddress]{revtex4}

\usepackage{amsmath}
\usepackage{bm}
\usepackage{graphicx}
\usepackage{epsfig}

\def\bea{\begin{eqnarray}}
\def\eea{\end{eqnarray}}
\def\beq{\begin{equation}}
\def\eeq{\end{equation}}

\newcommand{\ro}{\mbox{{\boldmath
$\rho$}}}

\newcommand{\db}{{{\bf d}}}
\newcommand{\rb}{\mbox{{\bf
r}}}

\newcommand{\bb}{{{\bf b}}}

\newcommand{\E}{{{\bf E}}}

\newcommand{\Vb}{{{\bf V}}}

\newcommand{\B}{{{\bf B}}}

\def\lsim{\mathrel{\rlap{\lower4pt\hbox{\hskip1pt$\sim$}}
    \raise1pt\hbox{$<$}}}         
\def\gsim{\mathrel{\rlap{\lower4pt\hbox{\hskip1pt$\sim$}}
    \raise1pt\hbox{$>$}}}         

\newcommand{\landau}{L.D.~Landau Institute for Theoretical Physics,
        GSP-1, 117940, Kosygina Str. 2, 117334 Moscow, Russia}

\begin{document}
\title{
FLUCTUATIONS OF ELECTROMAGNETIC FIELDS IN HEAVY ION COLLISIONS}

\author{B.G.~Zakharov}\affiliation{\landau}


\begin{abstract}
We perform quantum calculations
of fluctuations of the electromagnetic fields in $AA$ collisions
at RHIC and LHC energies. 
We find that in the quantum picture the field fluctuations 
are much smaller than predictions 
of the classical Monte-Carlo simulation with the Woods-Saxon nuclear density.
\end{abstract}

\maketitle

Non-central $AA$ collisions at high energies
can generate a very strong  magnetic field perpendicular 
to the reaction plane \cite{Kharzeev_B1,Toneev_B1}. 
In this talk I present results of quantum calculations
of fluctuations of the electromagnetic fields in $AA$ collisions
at RHIC and LHC energies based on the 
fluctuation-dissipation theorem (FDT) \cite{Callen}.
This issue is very important in the context of
the chiral magnetic effect and charge separation 
\cite{Kharzeev_B1,Kharzeev_CME_rev,Kharzeev_report} in
$AA$ collisions  
because the fluctuations
may partly destroy the correlation 
between the magnetic field direction and the reaction plane, and can lead 
to reduction of the $\B$-induced observables 
\cite{Liao_MC}.
Previously the field fluctuations 
have been addressed 
by Monte-Carlo (MC) simulation
with the Woods-Saxon (WS) nuclear distribution
using the classical Lienard-Weichert potentials
\cite{Skokov_MC,Deng_MC,Liao_MC}.
But the WS nuclear distribution ignores the collective 
quantum dynamics of the nuclear ground state.
The classical treatment of the electromagnetic field
may also be inadequate because, 
similarly to 
the van der Waals forces \cite{LL4}, 
it becomes invalid at large distances.

We consider the proper time region
$\tau\sim 0.2-1$ fm which is of the most interest for
the $\B$-induced effects in the quark-gluon plasma (QGP). 
We ignore the electromagnetic fields generated
by the induced currents in the QGP fireball 
after interaction of 
the 
colliding nuclei 
\cite{Z_maxw}.
We consider 
the right moving and left moving nuclei 
with velocities
$\Vb_{R}=(0,0,V)$ and $\Vb_{L}=(0,0,-V)$, and with the impact parameters
$\bb_R=(-b/2,0,0)$ and $\bb_L=(b/2,0,0)$. 
We take $z_{R,L}=\pm Vt$.
For each nucleus the electromagnetic field 
is a sum of the mean field and the fluctuating field
\beq
F^{\mu\nu}=\langle F^{\mu\nu}\rangle+\delta F^{\mu\nu}\,.
\label{eq:10}
\eeq
The mean fields $\langle \E\rangle$ and $\langle \B\rangle$  are given 
by the Lorentz transformation of the Coulomb field in 
the nucleus rest frame.
For two colliding nuclei 
the mean magnetic field at $\rb=0$  has only $y$-component. 
At $t\gg R_{A}/\gamma$ 
(here $\gamma=1/\sqrt{1-V^{2}}$ is the Lorentz factor, $R_A$ is the nucleus 
radius)   
in the region $\rho\ll t\gamma$ $\langle B_{y}(t,\ro,z=0)\rangle$      
takes a simple $\rho$-independent form 
\beq
\langle B_{y}(t,\ro,z=0)\rangle \approx
Zeb/\gamma^{2}t^{3}\,.
\label{eq:70}
\eeq

The contribution of each nucleus to the correlators of the electromagnetic
fields in the lab-frame may be expressed  via 
the correlators in the nucleus rest frame.
For $\gamma\gg 1$ the dominating fluctuations
in the lab-frame are the ones of the transverse  fields.
The transverse components  of the correlators 
of the electric and magnetic fields
can be written as 
\beq
\hspace{-.05cm}\langle \delta E_i \delta E_k \rangle =
\gamma^2\left[\langle \delta E_i \delta E_k \rangle+
V^2e_{3il}e_{3kj}\langle \delta B_l \delta B_j \rangle
\right]_{rf},
\label{eq:80}
\eeq 
\beq
\hspace{-.06cm}
\langle \delta B_i \delta B_k \rangle =
\gamma^2\left[\langle \delta B_i \delta B_k \rangle+
V^2e_{3il}e_{3kj}\langle \delta E_l \delta E_j \rangle
\right]_{rf},
\label{eq:90}
\eeq 
where $i,k$ are the transverse indices and   
the subscript $rf$ on the right-hand side of (\ref{eq:80}), (\ref{eq:90})   
indicates that  the correlators are calculated in the nucleus rest frame.

In calculations of the rest frame correlators
$\langle \delta E_l \delta E_j \rangle$,
$\langle \delta B_i \delta B_k \rangle$
(hereafter we drop the subscript $rf$)
with the help of the FDT
we follow 
the formalism of \cite{LL9} (formulated in the gauge $\delta A^{0}=0$). 
It allows to relate the time Fourier component of the vector potential 
correlator
\bea
\langle \delta A_i(\rb_1)\delta A_k(\rb_2)\rangle_{\omega}=
\frac{1}{2}\int dt e^{i\omega t}
\langle \delta A_i(t,\rb_1)\delta A_k(0,\rb_2)
+\delta A_k(0,\rb_2)\delta A_i(t,\rb_1)\rangle\hspace{2cm}
\label{eq:100}
\eea
and that of the retarded Green's function
\bea
D_{ik}(\omega,\rb_1,\rb_2)=-i\int dt e^{i\omega t}
\theta(t)
\langle \delta A_i(t,\rb_1)\delta A_k(0,\rb_2)
-\delta A_k(0,\rb_2)A_i(t,\rb_1)
\rangle\,.\hspace{2cm}
\label{eq:110}
\eea
In the zero temperature limit the FDT relation between (\ref{eq:100}) 
and (\ref{eq:110}) reads \cite{LL9} 
\bea
\!\!\langle \delta A_i(\rb_1)\delta A_k(\rb_2)\rangle_{\omega}\!=\!
-\mbox{sign}(\omega)
\mbox{Im} D_{ik}(\omega,\rb_1,\rb_2).
\label{eq:120}
\eea
The time Fourier components of 
the electromagnetic field correlators in terms of that for the 
the vector potential correlator (\ref{eq:100}) are given by
\beq
\langle \delta E_i(\rb_1)\delta E_k(\rb_2)\rangle_{\omega}=
\omega^{2}\langle \delta A_i(\rb_1)\delta A_k(\rb_2)\rangle_{\omega}\,,
\label{eq:130}
\eeq
\beq
\hspace{-.18cm}
\langle \delta B_i(\rb_1)\delta B_k(\rb_2)\rangle_{\omega}=
\mbox{rot}^{(1)}_{il}
\mbox{rot}^{(2)}_{kj}
\langle \delta A_l(\rb_1)\delta A_j(\rb_2)\rangle_{\omega}.
\label{eq:140}
\eeq

In the time region of interest ($t \gsim 0.2$ fm 
in the lab-frame) 
for each nucleus
the distance between the observation
point 
and  the center of the nucleus (in its rest frame)
is much bigger than $R_A$.
It allows one to treat each nucleus
as a point like dipole described by the dipole polarizability 
$\alpha_{ik}(\omega)$.
The field fluctuations are described 
by correction to the retarded Green's function
proportional to the dipole polarizability \cite{LL9}.
The retarded Green's function coincides with the Green's function of 
Maxwell's equation \cite{LL9}.
For the point like dipole at $\rb=\rb_A$
the equation for the retarded Green's function reads
\bea
\left[\frac{\partial^2}{\partial x_i\partial_l}-\delta_{il}\triangle
-\delta_{il}\omega^2
-4\pi\omega^2\alpha_{il}(\omega)\delta(\rb-\rb_A)\right]
D_{lk}(\omega,\rb,\rb')
=-4\pi\delta_{ik}\delta(\rb-\rb')\,.\hspace{1cm}
\label{eq:170}
\eea
The correction to $D_{ik}$ due to $\alpha_{ik}$
reads \cite{LL9}
\bea
\Delta  D_{ik}(\omega,\rb_1,\rb_2)=-
\omega^2 D_{il}^{v}(\omega,\rb_1,\rb_A)\alpha_{lm}(\omega)
D_{mk}^{v}(\omega,\rb_A,\rb_2)\,,
\label{eq:180}
\eea
where $D_{ik}^{v}$ is the vacuum Green's function given by
\beq
D_{ik}^{v}(\omega,\rb_1,\rb_2)=\frac{e^{i\omega r}}{r}\left[
-\delta_{ik}
\left(1+\frac{i}{\omega r}
-\frac{1}{\omega^2 r^2}\right)
+
\frac{x_ix_k}{r^2}\left(1+\frac{3i}{\omega r}
-\frac{3}{\omega^2 r^2}\right)\right]
\label{eq:190}
\eeq
with $\rb=\rb_1-\rb_2$.

For spherical nuclei the polarizability tensor can be written
as $\alpha_{ik}(\omega)=\delta_{ik}\alpha(\omega)$. 
$\alpha(\omega)$ is an analytical function 
of $\omega$ in the upper half-plane \cite{LL4}. It satisfies the 
relation $\alpha^{*}(-\omega^{*})=\alpha(\omega)$ \cite{LL4}. It means that
on the upper imaginary axis $\alpha(\omega)$ is real.
Using this fact, 
one can express 
the rest frame field correlators 
$
\langle \delta E_i(t,\rb)\delta E_k(t,\rb)\rangle
$,
$
\langle \delta B_i(t,\rb)\delta B_k(t,\rb)\rangle
$
via integrals 
of the type
$
I_n=\int_{0}^{\infty} d\xi \xi^n e^{-\xi}\alpha\left(\frac{i\xi}{2
  r}\right)\,
$ with $n=0-4$ \cite{Z_hep}.

The function $\alpha(\omega)$ reads 
\cite{LL4}
\beq
\alpha(\omega)=\frac{1}{3}\sum_{s}\left[
\frac{|\langle 0|\db|s\rangle|^2 }
{\omega_{s0}-\omega -i\delta}+
\frac{|\langle 0|\db|s\rangle|^2 }
{\omega_{s0}+\omega +i\delta}\right]\,,
\label{eq:280}
\eeq
where 
$
\db=\left(eN\sum_p \rb_p-e Z\sum_n \rb_n\right)/A 
$
is the dipole operator.
At $\omega>0$ 
the imaginary part of $\alpha(\omega)$ is connected with 
the dipole photoabsorption cross section 
\beq
\sigma_{abs}(\omega)=4\pi\omega\mbox{Im}\alpha(\omega)\,.
\label{eq:300}
\eeq
For heavy nuclei the dipole strength 
is dominated by the giant dipole resonance (GDR)
\cite{Greiner}. 
It appears as a broad peak in $\sigma_{abs}$
at $\omega\sim 14$ MeV.
We parametrize the dipole polarizability
for $^{197}$Au and $^{208}$Pb nuclei
by a single GDR state   
\beq
\alpha(\omega)=
c 
\left[
\frac{1}{\omega_{10}-\omega -i\Gamma/2}+
\frac{1}{\omega_{10}+\omega +i\Gamma/2}
\right]\,.
\label{eq:310}
\eeq

By fitting the data on the photoabsorption
cross section from \cite{GDR_Au} for $^{197}$Au and from \cite{GDR_Pb} for 
$^{208}$Pb 
we obtained the following values of the
parameters: $\omega_{10}\approx 13.6$ MeV, $\Gamma\approx 4.38$ MeV, 
$c\approx 18.2$ GeV$^{-2}$ for $^{197}$Au, and
$\omega_{10}\approx 13.3$ MeV, $\Gamma\approx 3.72$ MeV, $c\approx 18.93$
Gev$^{-2}$ for $^{208}$Pb.
\begin{figure}[]
\begin{minipage}[h]{0.49\linewidth}
\begin{center}
\includegraphics[width=0.97\linewidth,clip=]{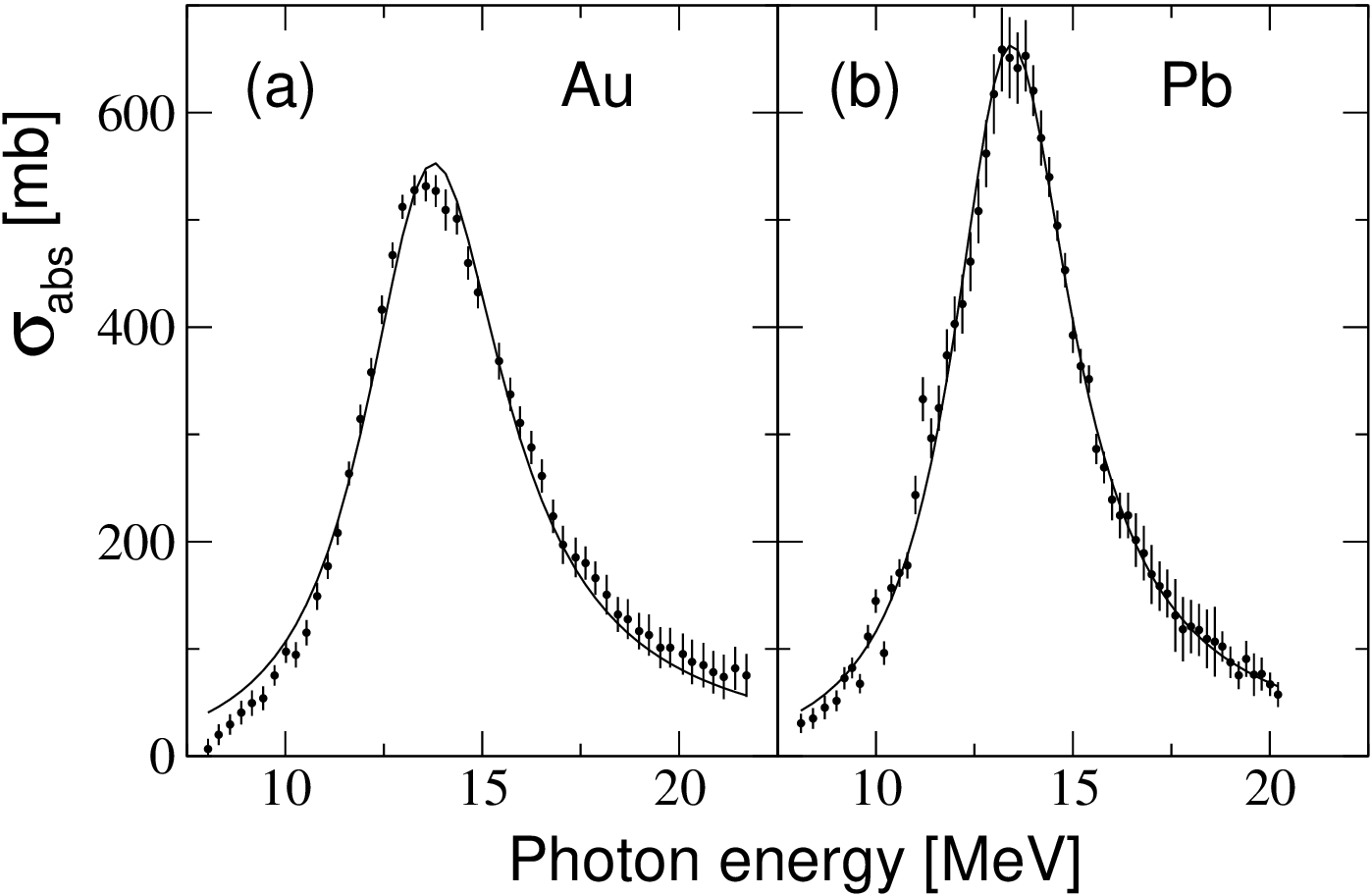}
\caption[.]
{Fit of the photoabsorption cross section in the GDR region 
to the experimental data for  $^{197}$Au and $^{208}$Pb targets.
The data are from Refs. \cite{GDR_Au} and \cite{GDR_Pb}, respectively.} 
\end{center}
\end{minipage}
\hfill
\begin{minipage}[h]{0.49\linewidth}
\begin{center}
\includegraphics[width=0.97\linewidth,clip=]{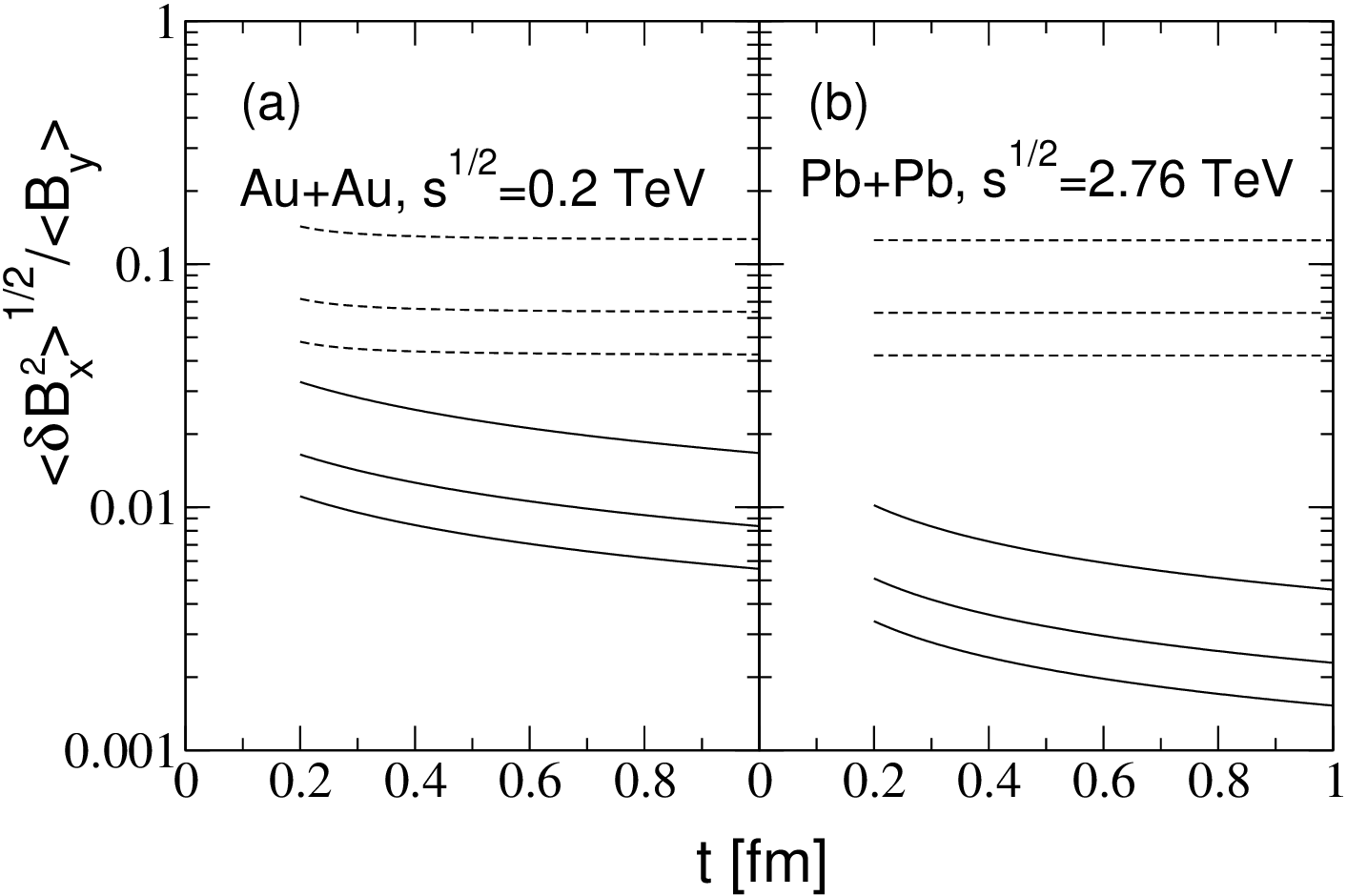}
\caption[.]
{The $t$-dependence of the ratio 
$\langle \delta B_x^2\rangle^{1/2}/\langle B_y\rangle$ 
at $\rb=0$ for Au+Au collisions at $\sqrt{s}=0.2$ TeV (left)
and for Pb+Pb collisions at $\sqrt{s}=2.76$ TeV
for the impact parameters $b=3$, $6$ and $9$ fm
(from top to bottom).
Solid lines are for quantum calculations, dashed
lines for classical MC calculations with
the WS nuclear density.}
\end{center}
\end{minipage}
\end{figure}
Fig.~1 illustrates the quality of our fit.
Using these parameters we calculated the fluctuations of the
nuclear dipole moment.
From (\ref{eq:280}), (\ref{eq:310}) 
one can obtain 
\beq
\langle 0 |\db^2|0\rangle=\frac{3}{\pi}
\int_{0}^{\infty}d\omega \mbox{Im}\alpha(\omega)
=\frac{6c}{\pi}
\mbox{arctg}\left(2\omega_{10}/\Gamma\right)\,.
\label{eq:320}
\eeq
This formula gives 
$\langle 0 |\db^2|0\rangle\approx 1.91$ fm$^2$
and $\langle 0 |\db^2|0\rangle\approx 2.02$ fm$^2$  for $^{197}$Au 
and $^{208}$Pb, respectively.
The classical MC calculation with the WS
nuclear density gives for these nuclei the values 
$\langle \db^2 \rangle\approx 9.89$ fm$^2$
and $\langle \db^2 \rangle\approx10.39$ fm$^2$.
Thus, we see that the classical
treatment overestimates the dipole moment squared by a factor of $\sim 5$.

At the 
center of the plasma fireball
the fluctuations of the direction of the magnetic field  are dominated
by the fluctuations of the component $B_x$ that vanishes without fluctuations.
In Fig.~2 we show our quantum and classical results for $t$-dependence 
of the ratio $\langle \delta B_x^2\rangle^{1/2}/\langle B_y\rangle$ 
at $x=y=0$ for several impact parameters
for Au+Au collisions at $\sqrt{s}=0.2$ TeV and Pb+Pb collisions at
$\sqrt{s}=2.76$ TeV.
This figure shows that the quantum treatment gives 
$\langle \delta B_x^2\rangle^{1/2}/\langle B_y\rangle$
smaller than
the classical one by a factor of $\sim 5-8$ for RHIC and 
by a factor of $\sim 13-27$ for
LHC. 
Thus, we see that in the quantum picture 
both for RHIC and LHC fluctuations of 
the direction of the magnetic field relative to the reaction plane should be 
very small.
Of course, experimentally 
the reaction plane itself cannot be
determined exactly. 
In the event-by-event measurements 
the orientation of the reaction 
plane is extracted from the elliptic flow in the particle
distribution
\cite{Ollitrault_v2,Voloshin_v2} 
(it is often called 
the participant plane), and it fluctuates around the real 
reaction plane. 
Calculations of the fluctuations
of the direction of the magnetic field relative to the 
participant
plane require a joint analysis of the field fluctuations
and of the fluctuations of the initial entropy deposition
that control the fluctuations of the orientation of 
the participant plane in the hydrodynamical
simulations of $AA$ collisions.
The initial entropy distribution is sensitive to the long range 
fluctuations of the nuclear density. 
Besides the nuclear fluctuations related to the GDR there are other collective 
nuclear modes 
\cite{Greiner} 
such as the giant monopole resonance 
and the giant quadrupole resonance that
may also be important for the participant plane fluctuations.
It would be of great interest to clarify the situation 
with the MC simulation 
with the WS nuclear density for these collective modes.
This is of great interest for the event-by-event hydrodynamic
simulations of $AA$ collision. 

In summary, 
we have performed a quantum
analysis of fluctuations of the electromagnetic field in $AA$ collisions
at RHIC and LHC energies. 
Our quantum calculations show that the field fluctuations are very small.
We have demonstrated that the classical picture overestimates strongly 
the field fluctuations.

This work is supported by the RScF grant 16-12-10151.

\end{document}